\documentclass[aps,prl,twocolumn,superscriptaddress]{revtex4-2} 
\usepackage{mathtools}
\usepackage{graphicx}
\usepackage[colorlinks=true,linkcolor=blue,citecolor=blue]{hyperref}
\usepackage{enumerate}
\usepackage{siunitx}
\usepackage{setspace}
\usepackage{float}

\begin{document}
\title{Narrowband four-photon states from spontaneous four-wave mixing}
\author{Yifan~Li}
\author{Justin Yu Xiang Peh}
\author{Chang Hoong Chow}
\author{Boon Long Ng}
\author{Vindhiya Prakash}
\affiliation{Centre for Quantum Technologies, National University of Singapore, 3 Science Drive 2, Singapore 117543}
\author{Christian~Kurtsiefer}
\affiliation{Centre for Quantum Technologies, National University of Singapore, 3 Science Drive 2, Singapore 117543}
\affiliation{Department of Physics, National University of Singapore, 2 Science Drive 3, Singapore 117551}
\email[]{christian.kurtsiefer@gmail.com}

\begin{abstract}
We observe time-correlated four photons within a correlation window of 20\,ns
from spontaneous four-wave mixing via a double-$\Lambda$ scheme in a cold cloud of $^{87}$Rb atoms. In contrast to high-power pulsed pumping of $\chi^{(2)}$ nonlinear processes in crystals, our scheme generates correlated four-photon states by direct continuous-wave pumping at nominal powers. We verify the presence of genuinely correlated four-photon states over accidentals by higher-order intensity cross-correlation measurements and accidental subtraction. We infer a time-correlated four-photon generation rate of $2.5(4) \times 10^6$ counts per second close to saturation. The photons produced are near-resonant with atomic transitions, and have a bandwidth in the order of MHz, making them readily compatible with quantum networking applications involving atoms. 
\end{abstract}

\maketitle

Multiphoton states, i.e. states with more than two photons entangled or correlated across single or multiple modes, are extremely useful resources for quantum sciences and technologies \cite{JWPan_Multiphoton}. At a foundational state, multiphoton Greenberger–Horne–Zeilinger (GHZ) states and W states have enabled powerful tests to disprove local realistic theories \cite{GHZOriginal,Pan2000quantum} and explore unique entanglement classes \cite{PhysRevLett.95.150404, Cirac2000}. 
Multiphoton states enable secure communication protocols \cite{Mark_SecretSharing, JWPan_Crypto} and also find application in quantum metrology \cite{Mitchell2004}. In the form of cluster states, they are essential for scalable and resource-efficient photonic quantum computing \cite{Raussendorf_OneWay, Browne_PRL}. States with four photons have also been used to encode decoherence-resistant quantum information \cite{Weinfurter_DecoherenceFree}. 

It is well known that some multiphoton states can be directly produced by strong pumping of non-linear processes like spontaneous parametric downconversion (SPDC), where the probability of producing more than one photon pair increases with the pump power \cite{dengGeneralizedTwomodeSqueezed1993, PhysRevA.100.041802}.  In directly pumping an SPDC process, the probability of producing entangled four photons is twice as high as producing two independent entangled pairs \cite{Weinfurter2001, Weinfurter2003}. Highly entangled W states have also been obtained from the higher-order component in a directly pumped SPDC process \cite{KieselWState}. 

Photons from SPDC typically have large bandwidths and correspondingly short coherence lengths, shorter than the length of the downconversion crystal itself.  Thus, in the above cases, pulsed pumps with large instantaneous powers and narrowband filters are often used to isolate and analyze correlated multiphoton states from SPDC. This leads to losses. Direct filter-free analysis of the rich temporal structure of higher-order correlated photons from SPDC \cite{bettelliCommentCoherenceMeasures2010} has been challenging due to the jitter and response averaging of detectors \cite{razaviCharacterizingHeraldedSinglephoton2009}.  The large bandwidth of photons from SPDC also limits their use in quantum memory and repeater schemes that require efficient interfacing with material quantum systems. 

Here, we demonstrate a bright source for narrowband time-correlated photon quadruplets, matched to atomic transitions, based on direct continuous-wave (cw) pumping of spontaneous four-wave mixing (SFWM) in a cold atomic cloud. In atomic clouds, SFWM is an excellent and bright \cite{Shiu2024, ChenPRR2022, Vuletic2006} alternative to SPDC for producing narrowband photon pairs with long coherence times \cite{duSubnaturalLinewidthBiphotons2008,DduNarrowbandBiphotonGeneration2008, kolchinElectromagneticallyinducedtransparencybasedPairedPhoton2007, srivathsanNarrowBandSource2013, balicGenerationPairedPhotons2005}. Photons from this process can be spectrally shaped to be narrower or wider than atomic transition linewidths, making them well-suited for quantum networking applications \cite{Kimble2008quantum}, such as memory, repeater \cite{sangouardQuantumRepeatersBased2011}, and entanglement distribution schemes involving atoms. Furthermore, their long coherence times, typically in the order of tens of nanoseconds, allow them to be well-resolved by off-the-shelf photon detection electronics.

We study correlated photon quadruplets from SFWM in a cold cloud of rubidium atoms using Hanbury Brown and Twiss (HBT) type setups, one in each of the correlated modes. We introduce an efficient technique for identifying three-fold and four-fold coincidences of the photons generated from the nonlinear interaction. We analyze the temporal distribution of the detected triplet and quadruplet coincidences, and observe that photon pairs bunch together in both measurements within a correlation window of 20\,ns. The aggregate coincidences detected within this window is significantly larger than the sum of accidentals detected at longer delays,
indicating a strong contribution from quadruplets correlated in time, over accidental/uncorrelated four-photon states.

\begin{figure}
    \centering
    \includegraphics[width=0.45\textwidth]{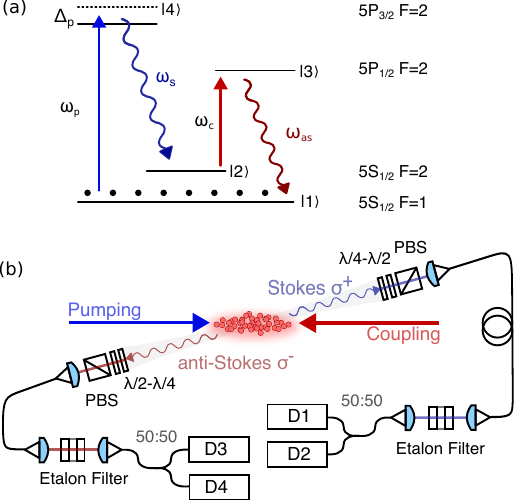}
   \caption{\label{fig:setup}(a) Energy levels involved in the Double-$\Lambda$ spontaneous four-wave mixing in $^{87}$Rb. Solid blue and red arrows indicate cw pump and coupling fields, respectively. Wiggly blue and red arrows indicate generated Stokes and anti-Stokes fields. Black dots indicate initialization of atoms in the $F=1$ hyperfine ground level. (b) Schematic of experimental setup. The pump and coupling beams have a waist of $\approx0.85$\,mm. The collection spatial mode is focused on the atomic ensemble with a waist of $175\,\mu\text{m}$. Detectors D1 and D2 detect the Stokes field, and D3 and D4 the anti-Stokes fields in a Hanbury-Brown and Twiss like setup. $\lambda/2$: half-wave plate, $\lambda/4$: quarter-wave plate, PBS: polarizing beamsplitter, D1-D4: single photon detectors. }
\end{figure}

In detail, our scheme is based on SFWM using a double-$\Lambda$ configuration of energy levels in a cold cloud of $^{87}$Rb atoms, similar to the systems reported in \cite{kolchinGenerationNarrowBandwidthPaired2006, kolchinElectromagneticallyinducedtransparencybasedPairedPhoton2007}.
The SFWM process is driven by a weak cw pump (of frequency $\omega_{p}$)
detuned by $\Delta_p$ from $|5{S}_{1/2}, F=1\rangle  \rightarrow |5{P}_{3/2},
F'=2\rangle $ and a strong cw coupling laser (of frequency $\omega_{c}$)
resonant to the $|5{S}_{1/2}, F=2\rangle  \rightarrow |5{P}_{1/2}, F'=2\rangle
$ transition. Nonlinear interaction of the pump and coupling fields with the
atomic medium generates correlated optical fields called Stokes and
anti-Stokes by convention. The Stokes photons are generated at a frequency
$\omega_s$ close to the $|5{P}_{3/2}, F'=2\rangle \rightarrow |5{S}_{1/2},
F=2\rangle$ transition and the anti-Stokes photons have a frequency
$\omega_{a}$ resonant to the $|5{P}_{1/2}, F'=2\rangle \rightarrow
|5{S}_{1/2}, F=1\rangle$ transition (see Fig.\ref{fig:setup}(a)).  The pump
and coupling fields are circularly polarized, orthogonal to each other, and
are directed at an elongated magneto-optical trap (MOT) of cold $^{87}$Rb
atoms, along the long axis in a counter-propagating configuration
(Fig.\ref{fig:setup} (b)). The SFWM process is precluded by initializing atoms
in the MOT into a state in the $|5S_{1/2}, F=1\rangle$ hyperfine ground level via optical pumping. The MOT cooling beams are switched off during the SFWM measurement. The optical depth (OD) of the atomic cloud is $\sim 30$. The spatial modes for collecting the Stokes and anti-Stokes photons are focused on the atomic ensemble with a waist of $175\ \mu\text{m}$. The collection modes form an angle of 1$^{\circ}$ with the pump and coupling fields to reduce background scattering. Polarization filters and temperature-controlled etalon filters (bandwidth $\approx$100\,MHz) are implemented in both Stokes and anti-Stokes collection arms to suppress unwanted photons. The photons collected in the Stokes and anti-Stokes arms are split using 50:50 fiber beamsplitters (BS) and detected using single photon detectors (D1-D4). A timestamp unit with 2\,ns timing resolution records the photon arrival times in each of these four detectors. Second, third and fourth-order field correlations are analyzed using this data.

A single frequency conversion process produces the following output state that can contain multiple Stokes and anti-Stokes photons \cite{Leonhardt, Lvovsky, sangouardQuantumRepeatersBased2011, deriedmattenTwoIndependentPhoton2004}:
\begin{equation}\label{eq:two_mode_squeezing_state2}
    \begin{split}
    |\Psi\rangle & =  \frac{1}{\beta} \sum_{n=0}^{\infty} \left( \alpha \right)^{n} |n,n \rangle  \, \\
    \end{split}
\end{equation}
 Here, $\beta\equiv \mathrm{cosh}\,\zeta$, $ \alpha \equiv
 \mathrm{tanh}\,\zeta $, $\zeta$ depends on the strength of the pump, the nonlinear interaction, and the duration of interaction, 
 and $|n,n\rangle$ indicates correlated Fock states with $n$ photons each in the Stokes and anti-Stokes modes. A complete expression for the interaction Hamiltonian and the nonlinear susceptibilities can be found in \cite{wenTransverseEffectsPairedphoton2006, DduNarrowbandBiphotonGeneration2008}. From Eq.~(\ref{eq:two_mode_squeezing_state2}), it is evident that at small interaction strengths ($\zeta\ll 1$) the probability of generating states with four photons ($P_4$) relates to the probability of producing pairs ($P_2$) as $P_4=P^2_2$. In this case the four-photon state corresponds to two Stokes and two anti-Stokes photons that are correlated and entangled, generated within a single SFWM process
 \cite{deriedmattenTwoIndependentPhoton2004}. Furthermore, this is twice the
 probability of four-photon states present if the output contains a Poissonian
 distribution of photons \cite{deriedmattenTwoIndependentPhoton2004} (Supplementary Material). 
 
 In the following, we characterize the composition of correlated quadruplets versus uncorrelated double pairs from our source using higher-order intensity correlation measurements and verify the presence of twice as many correlated quadruplets as uncorrelated four-photons.  We also compare rates of pair production versus four-photon production for varying pump powers.

\begin{figure}
    \centering
    \includegraphics[width=\linewidth]{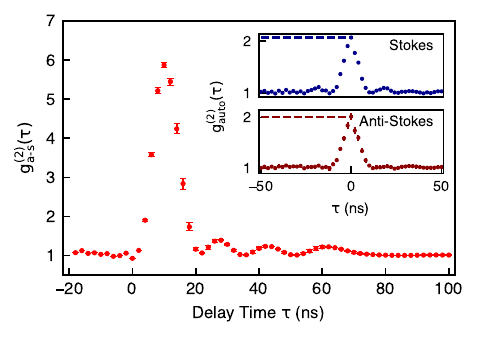}
    \caption{Normalized second-order correlation measurements. Main figure:
      Stokes---anti-Stokes cross-correlation as a histogram of coincidences
      for various detection delays $\tau$, normalized to the Stokes and anti-Stokes singles rates for a 2\,ns bin size and an integration time of 150\,s. Results averaged over 17 measurements. Oscillations are caused by the coupling field that drives the $|2\rangle \rightarrow |3\rangle$ transition at an effective Rabi frequency of $2\pi \times 55\, \text{MHz}\,$.  Insets: Unheralded autocorrelation measurements of Stokes photons (blue) with peak $g^{(2)}_{s,s}(0) = 2.07 \pm 0.02$  and anti-Stokes photons (red) with peak $g^{(2)}_{a,a}(0)=2.02 \pm 0.07$ (jointly labeled $g^{(2)}_{\text{auto}}(\tau)$). }
    \label{fig:crosscorrelation_w_auto}
\end{figure}

We measure the second-order intensity correlations as a first step towards characterizing the statistical properties of the generated fields. The normalized second-order correlation between stationary fields $\hat{E}_i$ in mode $i$, detected at time $t_i$, and $\hat{E}_j$ detected at time $t_j = t_i+\tau_{ji}$ is \cite{glauberQuantumTheoryOptical1963}  
\begin{equation}\label{eq:normg2}
    g^{(2)}_{ji}(\tau_{ji})= \frac{\langle\hat{E}^{\dagger}_{i}(t_{i})\hat{E}^{\dagger}_{j}(t_{i}+\tau_{ji})\hat{E}_{j}(t_{i}+\tau_{ji})\hat{E}_{i}(t_{i})\rangle}{\langle \hat{E}^{\dagger}_{j}(t_{i}+\tau_{ji})\hat{E}_{j}(t_{i}+\tau_{ji})\rangle \langle \hat{E}^{\dagger}_{i}(t_{i})\hat{E}_{i}(t_{i})\rangle }\ , 
\end{equation}
where $ i,j \in\{s, a\}$ for the Stokes ($s$) and anti-Stokes ($a$) modes.

The second-order autocorrelations $g^{(2)}_{{s,s}}(\tau)$, $g^{(2)}_{{a,a}}(\tau)$ and cross-correlation $g^{(2)}_{{s,a}}(\tau)$ were measured for pump and coupling powers of about $800\,\mu\text{W}$ and $10\,\text{mW}$, respectively, and a pump detuning of $\Delta_{p} = 2\pi \times 40\,\text{MHz}$. Results are shown in Fig.~\ref{fig:crosscorrelation_w_auto}. We infer a Stokes-anti-Stokes two photon correlation time of around $16$\,ns from the time constant of an exponentially decaying fit to the $g^{(2)}_{{s,a}}(\tau)$ results. 
The Stokes and anti-Stokes modes independently display thermal statistics as seen from their intensity autocorrelation at $\tau=0$ (inset in Fig.~\ref{fig:crosscorrelation_w_auto}).

\begin{figure}
    \centering
    \includegraphics[width=\linewidth]{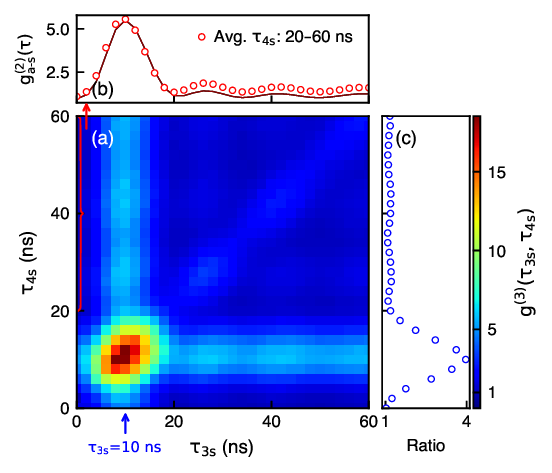}
    \caption{ Normalized third-order correlation. (a) Normalized triple coincidences $g^{(3)}_{a,a,s}$ for various delays $\tau_{3s}$ and $\tau_{4s}$ between a detection in D3 and D4,respectively, and a heralding Stokes photon in either of D1 or D2. Coincidences analyzed from data acquired over a measurement duration $T_m$ of 0.7\,h, normalized by the accidental triplet rate $R_{s}R_{3}R_{4}\delta t^{2}T_{m}$, where the time bin $\delta t =2\, \text{ns}$, and $R_i$ is the single count in detector D$i$. The $g^{(3)}_{a,a,s}$ peak value of 18 indicates strongly correlated triplets.  (b) Comparison of the vertical ridge with $g^{(2)}_{a,s}$. Red dots: $g^{(3)}_{a,a,s}$ results averaged over $\tau_{4s}$ from 20\,ns to 60\,ns.  Solid line: normalized cross-correlation $g^{(2)}_{a,s}(\tau_{3s})$ between a Stokes detection in D1 or D2 and an anti-Stokes detection in D3. (c) Peak to ridge ratio. Blue dots: Trace at $\tau_{3s} = 10\,\mathrm{ns}$, normalized by average value at $\tau_{3s} = 10$\,ns and $\tau_{4s  }= 20$ to $60$\,ns, i.e., at long delays. The peak is close to 4 times the mean value in the ridge.}
    \label{fig:triplet_w_corr}
\end{figure}

We analyze the temporal distribution of coincidences involving more than two detections to determine the ratio of correlated quadruplets to two independent pairs detected together by chance.
Since there is no physical mechanism that generates states involving only three photons, a measurement of triplet coincidences involving two Stokes and one anti-Stokes photons or two anti-Stokes and one Stokes photon provides similar information to a four-fold coincidence measurement of two anti-Stokes and two Stokes photons, while being faster to acquire and simpler to visualize.

The normalized third-order correlation between the Stokes and anti-Stokes modes from two anti-Stokes detections at times $t_3$ and $t_4$ and a Stokes detection at time $t_s$ is 
\begin{equation} \label{eq:normg3}
    \begin{aligned}
        & g^{(3)}_{a,a,s}(t_3,t_4,t_s) =  \\ & \frac{ \langle \hat{E}^{\dagger}_{s}(t_{\text{s}})\hat{E}^{\dagger}_{a}(t_{\text{3}}) \hat{E}^{\dagger}_{a}(t_{4})\hat{E}_{a}(t_{4})\hat{E}_{a}(t_{\text{3}})\hat{E}_{s}(t_{s}) \rangle}{\langle \hat{E}^{\dagger}_{s}(t_{s})\hat{E}_{s}(t_{s}) \rangle \langle \hat{E}^{\dagger}_{a}(t_{3})\hat{E}_{a}(t_{3}) \rangle\langle \hat{E}^{\dagger}_{a}(t_{4})\hat{E}_{a}(t_{4}) \rangle}\ ,
    \end{aligned}
\end{equation}
where the numerator leadsto the triple-coincidence rate $G^{(3)}_{a,a,s}(t_3,t_4,t_s)$. This can be expressed in terms of second-order correlations as shown in Eq.~(7) in the Supplementary Material. 

Figure~\ref{fig:triplet_w_corr} shows $g^{(3)}_{a,a,s}$ for triplets from an anti-Stokes detection each in D3 (at $t_3$) and D4 (at $t_4$), and a Stokes detection in either of D2 or D1 (at $t_s$), where the measurement was performed under the same conditions as the second order correlation measurements 
The results are represented in terms of relative delays $\tau_{3s}= t_3-t_s$ and  $\tau_{4s}= t_4-t_s$. The technique used to identify triplets from pair coincidences is described in the Supplementary Material. 

The features in Fig.~\ref{fig:triplet_w_corr} can be understood intuitively or by analyzing Eq.~(7) in the Supplementary Material 
over various delays. Given a coherence time $\Delta t$ for the Stokes and anti-Stokes photons, when $\tau_{3s}, \tau_{4s}, \tau_{34} \gg \Delta t$, the triplet rate reduces to the background accidental rate which is normalized to 1 in Fig~\ref{fig:triplet_w_corr}.
When $\tau_{3s}, \tau_{4s}, \gg \Delta t$ and $\tau_{34} \lesssim \Delta t$, the autocorrelation in the anti-Stokes mode dominates the result ($g^{(3)}_{a,a,s}(\tau_{3s},\tau_{4s},\tau_{34}) \rightarrow g^{(2)}_{a,a}(\tau_{34})$). In this case, the triplets are caused by the combination of an accidental click in the Stokes mode with a bunched thermal state in the anti-Stokes mode, forming the moderately bright diagonal in Fig.~\ref{fig:triplet_w_corr}. 

When $\tau_{3s}, \tau_{34} \gg \Delta t$ but $\tau_{4s} \lesssim \Delta t$ (horizontal ridge), or when $\tau_{4s}, \tau_{34} \gg \Delta t\,$ but $ \tau_{3s} \lesssim \Delta t$ (vertical ridge), the strong cross-correlation between anti-Stokes (in D4 or D3, respectively) and Stokes photon pairs are the dominant contributions. Here, the triplets are formed by a combination of a correlated Stokes-anti-Stokes pair with an uncorrelated additional photon in the anti-Stokes mode. Thus, the maximum mean value in the horizontal and vertical ridges at long delays is equal to $g^{(2)}_{s, a}(0)$ as seen in Fig.~\ref{fig:triplet_w_corr}(b). 

In the region where $\tau_{3s}, \tau_{34}, \tau_{4s} \lesssim \Delta t$ the
coincidences increase several-fold. Theoretically, the peak is expected to be
4 times the average in either the horizontal or vertical ridges (at delays
longer than $\Delta t$) when the output contains highly correlated four-photon
states (refer to Supplementary Material). We see from Fig.\ref{fig:triplet_w_corr} (c) that in our measurement, the $g^{(3)}_{a,a,s}$ peak is about four times the mean along the vertical ridge (outside the central 20\,ns window). Thus, we are confident that the output of the SFWM process contains twice as many strongly correlated four-photons as uncorrelated double-pairs, that contribute to the high three-fold coincidences in the triplet measurement.

\begin{figure}
    \centering
    \includegraphics[width=\linewidth]{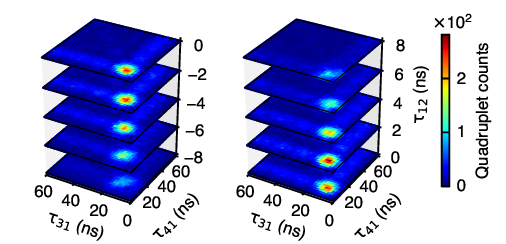}
    \caption{Quadruple-coincidence detection. Each slice shows unnormalized four-fold coincidences from a detection in each of D1 to D4 for a fixed delay $\tau_{12}$ with a 2\,ns time bin, and a range of delays $\tau_{31}$ and $\tau_{41}$. Data acquired over a measurement duration of 0.7\,h. The coincidences are peaked for $\tau_{12}=0\pm 2$\,ns and $\tau_{31}$ and $\tau_{41}= 8 \pm 2$\,ns.   }
    \label{fig:sheetstack}
\end{figure}

We search for four-fold coincidences between detections of two photons in the anti-Stokes mode and two photons in the Stokes mode for further analysis of four-photon states produced from our SFWM source.   
The normalized fourth-order cross-correlation is 
\begin{small}
\begin{equation}\label{eq:g4}
\begin{aligned}
 & {   g^{(4)}_{s,s,a,a}(t_1, t_2, t_3, t_4) =} \\
 & { \frac{\langle \hat{E}^{\dagger}_{a}(t_{\text{4}})
 \hat{E}^{\dagger}_{a}(t_{\text{3}}) \hat{E}^{\dagger}_{s}(t_{2})\hat{E}^{\dagger}_{s}(t_{1})\hat{E}_{s}(t_{1})\hat{E}_{s}(t_{2})\hat{E}_{a}(t_{\text{3}})\hat{E}_{a}(t_{\text{4}}) \rangle}{\langle \hat{E}^{\dagger}_{s}(t_{1})\hat{E}_{s}(t_{1}) \rangle \langle \hat{E}^{\dagger}_{s}(t_{2})\hat{E}_{s}(t_{2}) \rangle \langle \hat{E}^{\dagger}_{a}(t_{3})\hat{E}_{a}(t_{3}) \rangle\langle \hat{E}^{\dagger}_{a}(t_{4})\hat{E}_{a}(t_{4}) \rangle}},
\end{aligned}
\end{equation}
\end{small}where the numerator gives $G^{(4)}_{s,s,a,a}(t_1, t_2, t_3, t_4)$: the quadruplet rate for coincidences from two Stokes detections at times $t_1$ and $t_2$ respectively and two anti-Stokes detections at time $t_3$ and $t_4$, respectively.

We identify four-fold coincidences for detections at times $t_1$ to $t_4$ in
detectors D1 to D4, under the same experimental conditions as used in previous
measurements, with maximum delays up to 60\,ns. We represent the data as
sliced three dimensional histogram plots (see Fig.~\ref{fig:sheetstack}), where each slice shows quadruplets for a fixed delay $\tau_{12}$ and various relative delays $\tau_{31}$ and $\tau_{41}$. We see the maximum density of quadruplets clustered around $\tau_{12}= 0 \pm 2$\,ns and $\tau_{31}$ and $\tau_{41}= 8 \pm 2$\,ns. Outside a 20\,ns window centered at $(\tau_{12},\tau_{31},\tau_{41})=(0 \, \textrm{ns},10 \, \textrm{ns},10 \, \textrm{ns})$  the quadruplet count drops significantly, indicating the presence of highly-correlated quadruplets within 20\,ns. The horizontal and vertical ridges in slices as $\tau_{12}$ approaches $0$ arises from four-fold coincidences between accidentals and a correlated pair between D4-D1 or D3-D1, respectively. A relatively dull diagonal due to four-fold coincidences between accidentals and thermally bunched photons in D3-D4 can also be seen.

Due to the long coherence time of the Stokes and anti-Stokes photons, our triplet and quadruplet measurements are not limited by averaging effects due to detector resolution, which would have otherwise reduced the maximum of the triple and quadruple-coincidence peaks. 

\begin{figure}
    \centering
    \includegraphics[width=0.95\linewidth]{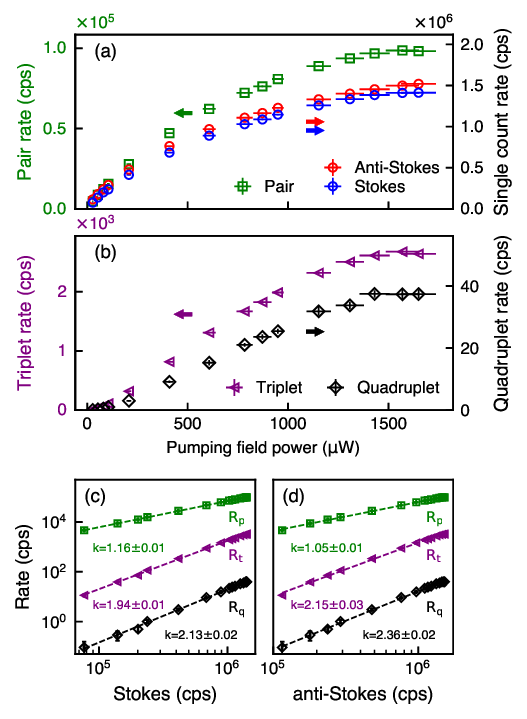}
    \caption{Top: Detection rates as a function of pump power. (a) Single count rates (right axis) of Stokes (blue circles) and anti-Stokes photons (red circles), and photon pair rate (green squares, left axis) as functions of pump power. (b) Total photon triplet rates (magenta triangles, left axis) and photon quadruplet rate (black diamonds, right axis) as functions of pump power. Bottom: Ratio of pairs, triplets and quadruplets to singles. The photon pair rate $R_{p}$ (green squares), photon triplet rate $R_{t}$ (magenta triangles), and photon quadruplet rate $R_{q}$ (black diamonds) from data in plots (a) and (b) represented in log-scale relative to the singles count rate $R_{\text{s}}$ in Stokes mode (c) and single count rate the anti-Stokes $R_{\text{a}}$ (d).  The change in single count rates is achieved by varying the pump power while keeping all other parameters constant. The detuning of the pump is 40\,MHz, while the coupling field is resonant with a fixed power of 10\,mW. The atomic cloud has an OD\,$\approx$\,30.}
    \label{fig:rate_vs_pumping_power}
\end{figure}
Next, we examine the pump power dependency of multiphoton states. Fig.~\ref{fig:rate_vs_pumping_power} (a), (b) show the total rates of Stokes/anti-Stokes singles ($R_{s/a}$), pairs ($R_{p}$), triples ($R_{t}$), and quadruples ($R_{q}$), aggregated over all detector combinations, defined within a coincidence window of $t_{c}=20\, \text{ns}$ and without subtraction of accidentals. Choosing $t_c$ of 20\,ns is appropriate as the detection of correlated double-pairs is peaked within this window as seen from the correlation measurements shown in Fig.~\ref{fig:crosscorrelation_w_auto}.

A clear relationship between the pairs and quadruplets for increasing pump powers is better visualized in Fig.~\ref{fig:rate_vs_pumping_power} (c) and (d), where the $R_p$, $R_t$ and $R_q$ are shown relative to $R_s$ and $R_a$, with axes in log scale. $R_{p}$ scales approximately linearly with $R_{s}$ and $R_{a}.$ 
The slopes of $R_t$ and $R_q$ are both close to 2, which is expected from the fact that triplet and quadruplet photons originate from the same physical processes. Furthermore, these measurements indicate that the rate at which double pairs are detected scales close to quadratically with the rate of detecting photon-pairs. This confirms that the photons in the double pairs are produced from a higher-order process in frequency conversion. 
At a pump power of $800\,\mu$W, which is close to saturation, we detect pairs at the rate of $7.1(3)\times 10^{4} \mathrm{\, cps}$ and quadruplets at the rate of 21(3),cps. We perform accidental subtraction obtain a detection rate of 3(1)\,cps for truly correlated quadruplets. Based on the detection rates and characterization of optical losses, we infer a quadruple generation rate of $ g_{q}=2.5 (4) \times 10^{6}\ \text{cps}$ and a pair generation rate of $g_{p}=1.3(3) \times 10^{7}\ \text{cps}$ at this power. Details on the procedure for accidental subtraction, channel losses and inferring generation rate can be found in the Supplementary Material. 

{\it Summary} -- So far, narrowband multiphoton states have been demonstrated by spatially multiplexing two SFWM processes, spontaneous Raman events, and cascaded geometries \cite{dongExperimentalRealizationNarrowband2017, parkGenerationBrightFourphoton2022, Wu_2016, hubelDirectGenerationPhoton2010}. Our results show that direct pumping holds the potential to be a simpler alternative to producing multiphoton states. Furthermore, the correlated four photons generated here may potentially also be time-energy entangled \cite{Park_17, deriedmattenTwoIndependentPhoton2004}. To our knowledge, microscopic models of SFWM as a collective process involving individual atomic emitters, have dealt only with the generation of photon-pairs \cite{raymondooiCorrelationPhotonPairs2007,jiangQuantumLangevinTheory2023}. Our results pave the way for the extension of such models to better understand the microscopic origin of correlated multiphoton states, i.e., the physical origin of enhancement in the generation of a correlated second pair in the same mode as an initially generated pair. In the case of correlated four-photons generated from pulsed pumping of SPDC, such enhancement has been attributed to the bosonic nature of photons that leads to preferential bunching of otherwise indistinguishable photons into the same mode \cite{deriedmattenTwoIndependentPhoton2004, Weinfurter2001, Weinfurter2003}. While this may explain our observations, there may be other contributions as well. The long coherence length of the generated photon-pairs in our experiment exceeds the length of the nonlinear medium and may potentially lead to stimulation of other pairs into the same mode, a process similar to the observations in \cite{Antia}. We invite theoretical analyses of our experiment to explore the above processes and identify the underlying mechanism. 

\begin{acknowledgments}
  This research was supported by the National Research Foundation, Singapore
  and A*STAR under its CQT Bridging Grant, and through the Ministry of
  Education, Singapore through grant MOET32024-0009.
\end{acknowledgments}

\bibliographystyle{apsrev4-2}
\nocite{}
%



\end{document}


\title{Supplementary Material for \\ Narrowband four-photon states from spontaneous four-wave mixing}
	\author{Yifan~Li}
	\author{Justin Yu Xiang Peh}
	\author{Chang Hoong Chow}
	\author{Boon Long Ng}
	\author{Vindhiya Prakash}
	\affiliation{Centre for Quantum Technologies, National University of Singapore, 3 Science Drive 2, Singapore 117543}
	\author{Christian~Kurtsiefer}
	\affiliation{Centre for Quantum Technologies, National University of Singapore, 3 Science Drive 2, Singapore 117543}
	\affiliation{Department of Physics, National University of Singapore, 2 Science Drive 3, Singapore 117551}
	\email[]{christian.kurtsiefer@gmail.com}
	
\maketitle

\appendix

\section{Event searching algorithms}\label{app:algorithms}
The search for multiple coincidences over timestamp data from four detectors is a computationally resource intensive task. We employ the following strategy to identify triplets and quadruplets. We first identify pair coincidences at various relative delays for each pair from the following possible pairs of Stokes-anti-Stokes detections  D1-D3, D1-D4, D2-D3, and D2-D4. 
The triplet coincidences are then identified based on the pair detections that share a photon arrival timestamp within a (short) coinciden time window. Quadruplet events are identified from triplet detections that share a common timestamp with a pair detection (again within an adequate coinciden time window).

For example, to identify triplet events involving D1, D3, and D4, we compare the timestamps of pairs between D1 and D3 detected at timestamps $t_{1}$ and $t_{3}$ with pairs between D1 and D4 detected at timestamps $t_{1}'$ and $t_{4}$. Pair events that share a common timestamp in D1 i.e. $t_{1}=t_{1}'$, are taken to form the triplet event $(t_{1}, t_{3}, t_{4})$. This information can be used to identify quadruplets detected across the four detectors. For this we compare pair events  between D2 (at $t_2$) and D4 (at time $t_4^*$)  with the previously identified triplet event with times $t_{1}, t_{3}, t_{4}$. When $t_{4}=t_{4}^*$, the events are combined to form a quadruplet detection with the timestamp $t_1, t_2, t_3,t_4$. Thus, we can efficiently search for pair, triplet, and quadruplet events from timestamp data and plot the temporal distribution of these coincidences.

\section{Correlated quadruplets vs uncorrelated double pairs}\label{app:corrvsuncorr}
We compare the four-photon component of the output (i.e., Stokes and
anti-Stokes photons) in two different scenarios: (a) when the output contains an aggregate of multiple independent SFWM events in the cloud with (b) when the output can be described as generated from a single coherent SFWM process. When a large number of independent SFWM processes take place, any $2n$ photon state can be satisfactorily described as $n$ independent pairs \cite{deriedmattenTwoIndependentPhoton2004}. In this case, the probability $P'_n$ of creating $n$ pairs within a certain time window is described by a Poisson distribution of mean $\mu$, $P'_n=e^{-\mu}\mu^n/n!$. For small $\mu$, $P'_4 \approx P'^2_2/2$. Here the four-photon states can be understood as a consequence of independent pairs that are coincidental in time.  However, if the output is the consequence of a single coherent SFWM process, it is described by Eq.~(1) in the main manuscript. At small interaction strengths ($\zeta\ll 1$) the probability of generating states with four photons ($P_4$) relates to the probability of producing pairs ($P_2$) as $P_4=P^2_2$. Here the quadruplets are time-correlated. The probability of generating four-photon states in the latter scenario, is twice as much as in the former scenario. 

The factor of two in the ratio between correlated four-photons and uncorrelated double pairs also manifests in a measurement of the third-order correlation between the modes as seen in Fig.~3 in the manuscript. From Eq.~(\ref{eq:TripletReduced}), 
when all delays are less than the coherence time ($\Delta t$) of the photons i.e., $\tau_{3s}, \tau_{34}, \tau_{4s} \lesssim \Delta t$, the third-order correlation increases several-fold. Using $g^{(2)}_{a,a}(0)=2$ (thermal statistics in each of the two correlated modes), we get the limit $g^{(3)}_{a,a,s}(0,0,0) \leq 4  g^{(2)}_{a,s}(0)$.  The theoretical maximum possible value of $g^{(3)}_{a,a,s}(0,0,0) = 4  g^{(2)}_{a,s}(0)$ is obtained when $g^{(2)}_{a,s}(0)\gg 1 $, a condition true for highly non-classical pair sources. Thus, theoretically, the peak is expected to be 4 times the average in either the horizontal or vertical ridges, or 2 times the sum of horizontal and vertical ridges, when the output contains highly correlated four-photon states. 

\section{Expanded Second, Third and Fourth order correlation functions  }\label{app:coincidencerates}

Here we express the double, triple and quadruple coincidence rates in terms of the phase sensitive first-order cross-correlation,
\begin{equation}
	C(\tau_{ij}) = \langle \hat{E}_i(t+\tau_{ij}) \hat{E}_{j}(t)\rangle,
\end{equation}
and the first-order autocorrelation
\begin{equation}
	R(\tau_{ij}) = \langle \hat{E}^\dagger_i(t+\tau_{ij}) \hat{E}_j(t)\rangle.
\end{equation}
Here $i$ and $j$ can be $s$ and/or $a$ to represent the Stokes and/or anti-Stokes mode, or  $\{i,j\}\in\{1,2,3,4\}$ to represent one of the four detection channels. 
Note, $ R(0)$ is the pair generation rate. 

The second order intensity correlation between fields $\hat{E}_i$ and $\hat{E}_j$ in modes $i$ and $j$ is
\begin{equation}\label{eq:normg2}
	g^{(2)}_{ji}(\tau_{ji})= \frac{\langle\hat{E}^{\dagger}_{i}(t_{i})\hat{E}^{\dagger}_{j}(t_{i}+\tau_{ji})\hat{E}_{j}(t_{i}+\tau_{ji})\hat{E}_{i}(t_{i})\rangle}{\langle \hat{E}^{\dagger}_{j}(t_{i}+\tau_{ji})\hat{E}_{j}(t_{i}+\tau_{ji})\rangle \langle \hat{E}^{\dagger}_{i}(t_{i})\hat{E}_{i}(t_{i})\rangle }\ . 
\end{equation}
When the state under consideration is the output of a parametric photon-pair production process such as SPDC or SFWM, the Gaussian moment factoring theorem can be applied to the intensity correlations in Eq.~(\ref{eq:normg2}) to give the following expressions for the normalized intensity cross-correlation \cite{shapiroSemiclassicalQuantumBehavior1994}
\begin{equation}\label{eq:normcross}
	g^{(2)}_{s,a}(\tau)=1+\frac{|C(\tau_{s,a})|^2}{|R(0)|^2},
\end{equation}
and the normalized intensity autocorrelation
\begin{equation}\label{eq:normauto}
	g^{(2)}_{i,i}(\tau)=1+\frac{|R(\tau_{i,i})|^2}{|R(0)|^2}.
\end{equation}

The triple-coincidence rate $G^{(3)}_{a,a,s}$ for an two anti-Stoke detections, one each at times $t_3$ and $t_4$ in D3 and D4, respectively, and a single Stokes detection at time $t_s$ in either of D1 or D2
\begin{equation}
	\begin{aligned}
		G^{(3)}_{a,a,s}(t_3,& t_4,t_s)=\\ 
		& \langle \hat{E}^{\dagger}_{s}(t_{\text{s}})\hat{E}^{\dagger}_{a}(t_{\text{3}}) \hat{E}^{\dagger}_{a}(t_{4})\hat{E}_{a}(t_{4})\hat{E}_{a}(t_{\text{3}})\hat{E}_{s}(t_{s}) \rangle,
	\end{aligned}
\end{equation}
can be similarly expanded and expressed in terms of relative delays to give \cite{razaviCharacterizingHeraldedSinglephoton2009,bettelliCommentCoherenceMeasures2010}
\begin{equation}\label{eq:tripletrate}
	\begin{aligned}
		G^{(3)}_{a,a,s}(\tau_{3s},\tau_{4s},\tau_{34})  = \, &   R(0)[R(0)^2 + |R(\tau_{34})|^{2}] \\ 
		& + R(0)[|C(\tau_{3s})|^{2}+|C(\tau_{4s})|^{2}] \\
		& + 2\text{Re}\{C(\tau_{3s})C^{\ast}(\tau_{4s})R(\tau_{34})\}.
	\end{aligned}
\end{equation}
Thus the accidental triplet events are a sum of four contributions:
accidental singles detected in each of the three channels, a correlated pair
between the Stokes mode and D3 (D4) with an accidental in D4 (D3), and
thermally bunched photons in the anti-Stokes (causing coincidences in D3 and
D4) with an accidental single in the Stokes mode (D1 or D2). Using
Eq.~(\ref{eq:normcross} (and Eq.~(\ref{eq:normauto}) this can be rewritten as
\begin{equation}
	\begin{aligned}
		G^{(3)}_{a,a,s}(\tau_{3s},\tau_{4s},\tau_{34}) 	=\, & R(0)^3 g^{(2)}_{a, a}(\tau_{34})+ R(0)^3 g^{(2)}_{a, s}(\tau_{3s})\\ 
		&+R(0)^3g^{(2)}_{a, s}(\tau_{4s}) - 2R(0)^3  \\
		&  +2 R(0)^3 \sqrt{g^{(2)}_{a, s}(\tau_{3s})-1} \\ & \sqrt{{g^{(2)}_{a, s}(\tau_{4s})-1}}
		\times \sqrt{{g^{(2)}_{a, a}(\tau_{34})-1}}.
	\end{aligned}
\end{equation}
The normalized third-order correlation is then
\begin{equation}
	\begin{aligned}
		g^{(3)}_{a,a,s}(\tau_{3s},\tau_{4s},\tau_{34})=&\frac{G^{(3)}_{a,a,s}(\tau_{3s},\tau_{4s},\tau_{34})}{R(0)^3}\\
		=\, & g^{(2)}_{a, a}(\tau_{34})+ g^{(2)}_{a, s}(\tau_{3s})\\ 
		&+g^{(2)}_{a, s}(\tau_{4s}) - 2  \\
		&  +2 \sqrt{g^{(2)}_{a, s}(\tau_{3s})-1} \sqrt{{g^{(2)}_{a, s}(\tau_{4s})-1}} \\
		& \times \sqrt{{g^{(2)}_{a, a}(\tau_{34})-1}}.
	\end{aligned} \label{eq:TripletReduced}
\end{equation}

Similarly, the quadruplet rate for detecting two Stokes photons at times $t_1,$ and $t_2$ and two anti-Stokes photons at times $t_3$ and $t_4$,
\begin{small}
	\begin{equation}\label{eq:g4}
		\begin{aligned}
			& {   G^{(4)}_{s,s,a,a}(t_1, t_2, t_3, t_4) =} \\
			& { \langle \hat{E}^{\dagger}_{a}(t_{\text{4}})
					\hat{E}^{\dagger}_{a}(t_{\text{3}}) \hat{E}^{\dagger}_{s}(t_{2})\hat{E}^{\dagger}_{s}(t_{1})\hat{E}_{s}(t_{1})\hat{E}_{s}(t_{2})\hat{E}_{a}(t_{\text{3}})\hat{E}_{a}(t_{\text{4}}) \rangle}.
		\end{aligned}
	\end{equation}
\end{small}

Applying the Gaussian moment factoring theorem, we obtain the following expression in terms of relative delays. 
\begin{equation}\label{eq:quadrupletrate}
	\begin{aligned}
		G^{(4)}_{s,s,a,a}(\tau_{34}, \tau_{24}, & \tau_{14},\tau_{23} ,\tau_{21},\tau_{13})=R(0)^4 \\
		&+R(0)^2[ |R(\tau_{43})|^2+|R(\tau_{21})|^2]\\
		&+R(0)^2[ |C(\tau_{23})|^2+|C(\tau_{24})|^2] \\ 
		&+R(0)^2[|C(\tau_{13})|^2+|C(\tau_{14})|^2]\\
		&+2R(0)\mathrm{Re}\{R(\tau_{34})C^*(\tau_{24})C(\tau_{23})\}\\
		&+2R(0)\mathrm{Re}\{R(\tau_{34})C^*(\tau_{14})C(\tau_{13})\}\\
		&+2R(0)\mathrm{Re}\{R(\tau_{21})C^*(\tau_{14})C(\tau_{24})\}\\
		&+2R(0)\mathrm{Re}\{R(\tau_{21})C^*(\tau_{13})C(\tau_{23})\}\\
		&+ |R(\tau_{43})|^2|R(\tau_{21})|^2 \\ 
		&+|C(\tau_{13})|^2|C(\tau_{24})|^2+|C(\tau_{14})|^2|C(\tau_{23})|^2\\&+2\mathrm{Re}\{C^*(\tau_{24})C^*(\tau_{13})C(\tau_{14})C(\tau_{23})\}]\\
		&+2 \mathrm{Re}\{R(\tau_{34})R(\tau_{12})C^*(\tau_{24})C(\tau_{13})\}\\
		&+2 \mathrm{Re}\{R(\tau_{34})R(\tau_{21})C(\tau_{23})C^*(\tau_{14})\}\,.\\
	\end{aligned}
\end{equation}

From the 17 terms in Eq.~(\ref{eq:quadrupletrate}), all terms other than the last three are contributions due to accidentals. Terms 2-7 are due to two accidentals combined either with a correlated Stokes-anti-Stokes pair or with thermally bunched photons in either of the Stokes or anti-Stokes modes. Terms 8-11 are caused by an accidental combined with a correlated triplet in three detectors. Term 12 is from bunching in both the Stokes and anti-Stokes modes. Terms 13 and 14 are correlated pairs from separate SFWM events contributing to four-fold coincidences. Terms 15, 16 and 17 are due to correlated double-pairs from the same SFWM event.

\section{Accidental Correction Procedure}\label{app:accidental}
Correction must be performed to eliminate contributions from accidental coincidences between uncorrelated events from different channels. This can be visualized as the relative probability of independent events falling within the same coincidence time window $t_c$ (i.e. random chance). For coincidences between two detectors, the accidental rate is $ t_c R_i R_j $
for singles rates $R_i$ and $R_j$ in detectors D$i$ and D$j$, respectively. Excess coincidence events after correction can then only be attributed to actual correlations between channels: in the case of two detectors D$i$ and D$j$ with an observed pair rate of $R_{ij}$, the correlated pair rate $c_{ij}$ is given by
\begin{equation}
	c_{ij} = R_{ij} - t_c R_i R_j.
\end{equation}

The total correlated pair rate is $c_p=\sum_{i,j}c_{ij}$ for $\{i,j\} \in \{\{1,3\},\{1,4\},\{2,3\}, \{2,4\}\}$.

These correlated pairs (and separately, accidentals) can be modeled as a separate stream of events that factor into the calculation of higher-order accidentals, e.g. 3-fold accidentals between channels $i,j$ and $k$ occur due to accidental coincidences across the individual channels $\{R_i,R_j,R_k\}$, as well as pairs with the remaining channel $\{c_{ij},R_k\}$, $\{c_{ik},R_j\}$ and $\{c_{jk},R_i\}$. For instance, in Fig.~3 in the manuscript, these correspond to the general background horizontal, vertical ridges and diagonal ridges when $\{i,j,k\}=\{s,3,4\}$. 

The correlated triplet rate is thus
\begin{equation}\label{eq:tripletcorr}
c_{ijk} = R_{ijk} - t_c\left(c_{ij}R_{k} + c_{jk}R_i + c_{ik}R_j\right) - t_c^2R_{i}R_{j}R_{k}
\end{equation}
given an observed triplet rate of $R_{ijk}$.

It can be seen that each of the individual terms contributing to the $n$-fold coincidences correspond to a possible partitioning of the set of all channels, with the total number of partitions given by Bell's number $B_n$ (i.e. $B_2 = 2$, $B_3 = 5$, $B_4=15$). We write out explicitly the exhaustive 14-term correction performed for 4-fold coincidences,
\begin{equation}
	\begin{aligned}
	c_{1234} &= R_{1234} \\
	&- t_c\left(c_{12}c_{34} + c_{13}c_{24} + c_{14}c_{23}\right) \\
	&- t_c\left(c_{123}R_{4} + c_{124}R_3 + c_{134}R_2 + c_{234}R_1\right) \\
	&- t_c^2\left(c_{12}R_{3}R_{4} + c_{13}R_2R_4 + c_{14}R_2R_3\right) \\
	&- t_c^2\left(c_{23}R_1R_4 + c_{24}R_1R_3 + c_{34}R_1R_2\right) \\
	&- t_c^3R_{1}R_{2}R_{3}R_{4}.
	\end{aligned}
\end{equation}

We also make a small note that this correction slightly overestimates the actual accidental rate~\cite{janossy1944rate} due to the $n$-fold coincidence calculation method containing an implicit ordering of events that introduces excess accidentals. 

In the experiment, this overcompensation is minimized by using a small 20\,ns coincidence window. At fixed pump and coupling powers of $800\, \mu$W and 10\,mW, respectively, and $\Delta_p=40$ MHz, the mean value of singles rates for Stokes photons is $R_s=(1.04 \pm 0.07)\times 10^{6} \text{\ cps}$ and anti-Stokes photons $R_{a}= (1.10 \pm 0.06) \times 10^{6} \text{\ cps}$. Performing accidental subtraction as outlined above we obtained a correlated-pair detection rate $c_p= (4.8 \pm 0.3) \times 10^{4} \text{\ cps}$. 

As we can see from Eq.~(\ref{eq:tripletcorr}), the accidental corrected rate depends on the combination of detectors chosen. Thus, we report detector-specific accidental-corrected triplet rates. The corrected rate of detected photon triplets consisting of one Stokes photon and one anti-Stokes photon each in D3 and D4 is $c_{134}+c_{234}=(251 \pm 10)\text{\ cps}$, while the rate for triplets consisting of a Stokes photon each in D1 and D2 and one anti-Stokes photon is $ c_{123}+c_{124}=( 246 \pm 7)\text{\ cps}$.
The correlated photon quadruplet rate after accidental subtraction for a detection in each of the four detectors, is found to be 
$c_{q}=(2.9 \pm 0.4) \text{\ cps}$. 

We infer the generation rates of pairs $g_{p}$ and double-pairs $g_{q}$ from detected accidental-corrected rates of pairs, triplets and quadruplets by factoring in the losses as described in Appendix~\ref{app:generationrate}.
Most of the accidentals are dominated by coincidences between uncorrelated souble pairs (5.5\,cps), followed by coincidences from a pair and two singles (5.6\,cps) and coincidences from a triplet combined with a single (4.8\,cps).

\section{Channel Losses}\label{app:ChannelLoss}
We characterize the losses in each channel to estimate the rate of correlated
pairs and double-pairs directly generated from the SFWM process. The total
transmission and detection probability in each channel $k \in\{1,2,3,4\}$,
which includes transmission of the collection and filtering setup, splitting
efficiency of the fiber-based 50:50 beamsplitter, and quantum efficiency of
the detector in the respective channel is denoted by $\eta_{k}$. The optical
losses in each channel are determined by measuring the transmission of a laser
beam (at the target wavelength) from outside the vacuum chamber to just before
the detector in each channel. We measure transmissions of $\approx 11.5\%$
each for channels 1 and 2 (with detectors D1 and D2, respectively) and $
12.5\%$ each for channels 3 and 4 (corresponding to D3 and D4, respectively),
which include the contributions from the filter Etalon and fiber
beamsplitter. Including the quantum efficiencies of avalanche photodiodes used
as single photon detectors in each channel (about 60-70$\%$ per detector), the
measured efficiencies $(\eta'_k)$ are 0.078, 0.083, 0.080, and 0.067 for
channels 1,2,3, and 4, respectively. These values provide an estimate of the
upper bound for effective efficiencies, as they do not account for absorption
in the atomic ensemble or spatial mode mismatch between the photons and the
collection optics.

Since we expect additional losses that are frequency specific to the Stokes and anti-Stokes modes, we define $\eta_i = \eta_s\eta'_i$ for $i \in \{1,2\}$ and $\eta_j = \eta_{a}\eta'_j$ for $j \in \{3,4\}$. We then use equations \ref{eq:triplet_detection} and \ref{eq:four_detection} to estimate $\eta_s$ and $\eta_{a}$. We infer additional losses that are $1-\eta_s=19 \%$ for channels in the Stokes arm and $1-\eta_{a}=8 \%$ for the anti-Stokes channels, which we attribute to a combination of above mentioned factors. Taking into account these additional losses, the total efficiencies are $\eta_1=0.022$ and $\eta_2=0.023$  for channels 1 and 2 pertaining to the Stokes modes, and the total efficiencies are $\eta_3 = 0.025$ and $\eta_4 = 0.021$   for channels 3 and 4 of the anti-Stokes mode.

\section{Generation Rates from Detection Rates}\label{app:generationrate}

We infer the generation rates of pairs $g_{p}$ and correlated double-pairs $g_{q}$ from the detected accidental-corrected rates of pairs, triplets and quadruplets by factoring in the channel losses as follows. The accidental-corrected quadruplet rate $c_{q}$ is solely contributed to by generated correlated double-pairs. There are four possible combinations by which each of the two Stokes photons reach D1 and D2 and each of the two anti-Stokes photons reach D3 and D4. This results in
\begin{equation}\label{eq:four_detection}
	c_{q} = 4 g_{q} \eta_{1}\eta_{2}\eta_{3}\eta_{4}\,.
\end{equation}

Similarly, double-pair generations are the sole contributors to accidental-corrected triplet coincidences. A triplet between detectors D1, D3 and D4 occurs from two possible combinations by which the two anti-Stokes photons reach one of D3 and D4 each (leading to the factor $2 \eta_3 \eta_4$), combined with the probability that at least one of the two Stokes photons reaches D1 (leading to the factor $1- (1-\eta_1)^2$ where $(1-\eta_1)^2$ is the probability that neither of the two Stokes photons reaches D1). Applying this to all combinations of triplet detections, 
\begin{equation}\label{eq:triplet_detection}
	\begin{aligned}
		c_{123} &= 2g_{q} \eta_{1}\eta_{2}(2-\eta_{3})\eta_{3} \\
		c_{124} &= 2g_{q} \eta_{1}\eta_{2}(2-\eta_{4})\eta_{4} \\
		c_{134} &= 2g_{q} \eta_{3}\eta_{4}(2-\eta_{1})\eta_{1} \\
		c_{234} &= 2 g_{q} \eta_{3}\eta_{4}(2-\eta_{2})\eta_{2}\,. \\ 
	\end{aligned}
\end{equation}
Here $c_{ijk}$ is the rate of correlated triplets between detectors $i$, $j$ and $k$. We use this to estimate mean values of $g_{q}$. 

Since we have non-number-resolving detectors, both pairs and correlated
double-pairs from the SFWM process contribute to accidental-corrected pair
coincidences. A coincidence between D1 and D3 can be caused by a generated
pair where the Stokes photon is detected in D1 and the anti-Stokes is detected
in D3 ($\eta_1 \eta_3$) or the probability that at least one of two Stokes and
two anti-Stokes photons from a double-pair reach D1 and D3, respectively, ($
(1- (1-\eta_1)^2)(1- (1-\eta_3)^2)= \eta_1\eta_3(2-\eta_1)(2-\eta_3)$). This
results in
\begin{equation}\label{eq:pair_detection}
	\begin{aligned}
		c_{13} &= \eta_{1}\eta_{3}(g_{p} + g_{q} (2-\eta_{1})(2-\eta_{3})) \\
		c_{14} &=  \eta_{1}\eta_{4}(g_{p} +g_{q} (2-\eta_{1})(2-\eta_{4}))\\
		c_{23} &=  \eta_{2}\eta_{3}(g_{p} + g_{q}(2-\eta_{2})(2-\eta_{3}))\\
		c_{24} &=  \eta_{2}\eta_{4}(g_{p} +g_{q} (2-\eta_{2})(2-\eta_{4}))\,, \\
	\end{aligned}
\end{equation}
where $c_{ij}$ is the accidental-corrected pair rate between channels $i$ and $j$. 

We use use accidental-corrected pair, triplet and quadruplet rates to obtain values for $g_{p}$ and $g_{q}$, reported in the manuscript. 

\bibliographystyle{apsrev4-2}
%

